\documentclass[a4paper]{jpconf}
\usepackage{graphicx}
\usepackage[usenames,dvipsnames]{pstricks}

\newcommand{\be}[1]{\begin{equation}#1 \end{equation}}

\def\viceversa{{\it vice versa\/}}
\def\eg{{\it e.g.\ }}
\def\etal{{\it et al.\ }}
\def\ie{{\it i.e.\ }}

\begin{document}
\title{Critical phenomena and information geometry in black hole physics}

\author{Jan E. {\AA}man and Narit Pidokrajt}

\address{Department of Physics \\ Stockholm University \\ 106 91 Stockholm \\ Sweden }

\ead{ja@fysik.su.se, narit@fysik.su.se}

\begin{abstract}
We discuss the use of information geometry in black hole physics and present the outcomes. The type of information geometry we utilize in this approach is the thermodynamic (Ruppeiner) geometry defined on the state space of a given thermodynamic system in equilibrium. The Ruppeiner geometry can be used to analyze stability and critical phenomena in black hole physics with results consistent with those from the Poincar\'e stability analysis for black holes and black rings. Furthermore other physical phenomena are well encoded in the Ruppeiner metric such as the sign of specific heat and the extremality of the solutions. The black hole families we discuss in particular in this manuscript are  the Myers-Perry black holes. 

\end{abstract}

\section{Introduction}

In this talk we discuss the use of information geometry, in particular the \emph{thermodynamic geometry} \cite{ruppeiner79}, also known as Ruppeiner geometry, of various black hole (BH) families. This has been studied over the past few years in \eg \cite{Aman:2003ug}, \cite{Shen:2005nu},  \cite{Shen:2005nu}, \cite{Medved:2008es}, \cite{Chakraborty:2008zzd}, \cite{Sarkar:2008ji} and recently in \cite{thesis} among several dozens of papers devoted to the use of this method to study BHs. Our results so far have been physically suggestive \cite{Aman:2005xk}, particularly in the Myers-Perry (MP) Kerr BH case where the curvature singularities signal the initial onset of thermodynamic instability of such BH. The geometrical patterns are given by the curvature of the Ruppeiner metric\footnote{This metric is conformal to the so-called Weinhold  metric via $g^W_{ij} = T g^R_{ij}$ where $T$ is thermodynamic temperature of the system of interest.} defined as the Hessian of the entropy on the state space of the thermodynamic system 
\begin{equation}
g^R_{ij} = - \partial_i \partial_j S(M, N^a),
\end{equation}
where $M$ denotes mass (internal energy) and $N^a$ are other parameters such as charge and spin. The minus sign in the definition is due to concavity of the entropy function. Interpretations of the geometries associated with the metric are discussed in \cite{Ruppeiner:1995zz} plus references therein. It has been argued that the curvature scalar of the Ruppeiner metric measures the complexity of the underlying interactions of the system, \ie the metric is flat for the ideal gas whereas it has curvature singularities for the  Van der Waals gas.  We take note that most interesting Ruppeiner metrics that we encounter have curvature singularities which are physically suggestive, but there are some known flat Ruppeiner metrics that can be understood from a mathematical point of view \ie we proved in  {\AA}man  \etal \cite{Aman:2006kd}  a flatness theorem  which states that Riemann curvature tensor constructed out of the negative of the Hessian of the entropy of the form 
\begin{equation}
S = M^k f(Q/M)
\end{equation}
will vanish, where $f$ is an arbitrary analytic function and $k \neq 1$. The latter condition is necessary in order for the metric to be nondegenerate. This theorem has proven useful in our work on the dilaton BHs \cite{Aman:2007ae} as it allows us to see the local geometry already by glancing at the entropy function. We also note that the signature of the Ruppeiner metric corresponds to the sign of the system's specific heat, \ie the signature is Lorentzian for system with negative specific heat and Euclidean when all specific heats are positive.

We have observed that the Ruppeiner curvature scalar diverges in the extremal limits\footnote{\ie where the BH temperatures vanish.} of the Kerr, MP and Reissner-Nordstr\"om (RN) AdS BH whilst it is vanishing for RN, BTZ and dilaton BHs \cite{thesis}. Despite the flat thermodynamic geometry one can extract useful information on  the BH solutions by plotting the state space of such the flat geometry. Normally it is done by first transforming the metric into a manifestly flat form and then bringing it into a recognizably Minskowskian form.  
\begin{figure}
\centering
\includegraphics[scale=.5]{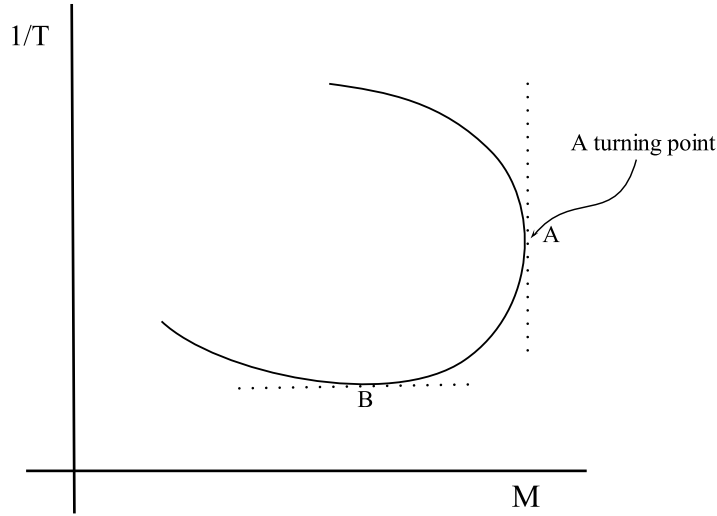}
\label{fig1}
\caption{\sl \small In the Poincar\'e method one plots the conjugacy diagram \ie the inverse temperature versus mass in our case.  The turning point according to the Poincar\'e method is where there is a vertical tangent to the curve, \ie point A. This is where a change of stability can take place. No change of stability happens at point B. }
\end{figure}

Incidentally there is a so-called {\it Poincar\'e}'s linear series method for analyzing stability in non-extensive systems. The simplicity of this method is owing to the fact that it utilizes only a few thermodynamic functions such as the fundamental relations in order to study/analyze (in)stabilities. This method can thus be applied to BHs although they are non-extensive systems. Non-extensitivity in BHs is due  to self-gravitation and furthermore the BHs cannot be subdivided and the BH entropy scales with area instead of the volume\footnote{This makes the issue of dynamical and thermodynamic (in)stabilities subtler. In extensive systems thermodynamic stability normally implies dynamical stability.}. For the analysis we use the recipes given in \cite{Arcioni:2004ww} without further elaboration. In the Poincar\'e method\footnote{The proof of this is given in \cite{katz} and \cite{sorkin}.} one plots a conjugacy diagram \eg an inverse temperature versus mass and we can infer some information about the existence of instability if the turning point is present (see Fig.~\ref{fig1}).

\section{Higher-dimensional black holes}

In higher dimensions BHs are more interesting objects due to richer rotation dynamics, and the appearance of extended black objects such as black strings. For more detailed and complete information see \eg  \cite{Emparan:2008eg}. The higher-dimensional generalization of the 4D BHs is the Myers-Perry BHs\footnote{The MP BHs can be regarded as the higher-dimensional versions of the Kerr solution.}. In 5D there is a black ring solution \cite{Emparan:2001wn}  which is a BH with a horizon topology $S^2 \times S^1$ in asymptotically flat spacetime.

\subsection*{Critical phenomena in higher-dimensional black holes}

Black strings and black branes suffer from instabilities known as the Gregory-Laflamme instability \cite{Gregory:1993vy}. The MP black holes with only one angular momentum turned on when spinning ultrafast also suffer from instabilities. The Ruppeiner curvature scalar~\cite{Aman:2005xk} for the MP BH with one spin is singular at 
\be{
4\frac{J^2}{S^2} = \frac{d-3}{d-5}.
}
which coincides with that found by Emparan and Myers in \cite{Emparan:2003sy} but in different coordinates\footnote{Note that this is for the 5D MP BH with only one spin turned on. In principle there can be two spins for the MP BH in 5D.} and the BH temperature reaches its minimum at this point. This is where the MP BH starts behaving like black membrane (which is an unstable object) but the instability is believed to set in somewhat later when the mass of the system decreases plus that there are unstable dynamical modes due to metric perturbations which the Poincar\'e method does not see such as those recently investigated (in various dimensions) by Dias \etal in \cite{Dias:2009iu}. Attempts to analyze such the instabilities by means of thermodynamic geometry are being made in \cite{inprogress1}. 

\begin{figure}
\centering
\includegraphics[scale=.5]{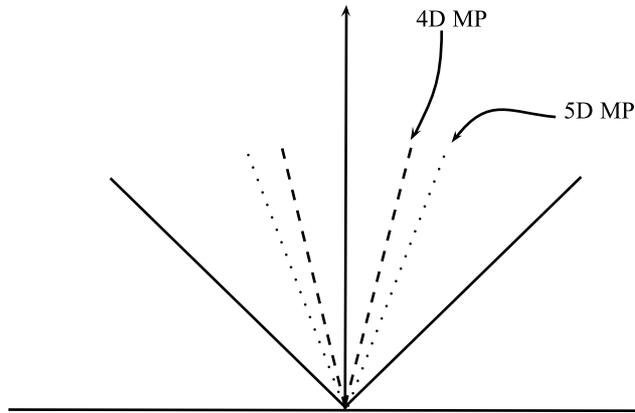}
\caption{\sl \small The state space of the 4D and 5D MP BHs. Note that as the number of dimensions increases the wider the opening angle becomes. The extremal limits lie on the edge of the wedge, and on the thermodynamic cone entropy is vanishing.}
\label{fig12}
\end{figure}

The Weinhold metric for the MP BH is flat and can be brought into a manifestly flat form by coordinate transformations. The state space of the MP BHs appears as a wedge embedded in a Minkowski space and we call this diagram a \emph{thermodynamic cone}. As mentioned above one can obtain some information from the thermodynamic cone \eg the BH entropy is vanishing on the light cone whereas the edge of the wedge is where $T = 0$ which is the BH's extremal limit. The opening angle is unique for each BH and it increases as the number of dimensions increases. As we increase the number of dimensions the opening angle tends to the right angle (see Fig.~\ref{fig12}). 

\subsection*{Black rings}

In \cite{Arcioni:2004ww} the authors show that the method of  thermodynamic geometry is consistent with the Poincar\'e stability analysis, and they are able to prove that one of the black ring\footnote{They consider large and small black rings categorized by the angular momentum the black ring possesses.} branches is always locally unstable, showing that there is a change of stability at the point where the two black ring branches meet. Their results using two different methods (Ruppeiner and Poincar\'e) are consistent with each other.

\section{Summary and outlook}

Information geometry is a new approach for studying BH thermodynamics and possibly BH instabilities. This approach opens up new perspectives and sheds light on critical phenomena in BH systems. The signature of the thermodynamic metric and the curvature singularities correspond to the sign of the specific heat and the extremal limit of the BHs respectively. The curvature singularity is physically suggestive in that it signals the initial onset of instability in higher dimensional BHs. This method has so far been consistent with the Poincar\'e analysis\footnote{We would like to point out that a divergence in Ruppeiner curvature scalar does not necessarily imply a turning point in the conjugacy diagram and \viceversa. We note also that one cannot in general anticipate to obtain all the information on dynamic stability from the Poincar\'e method but should the turning point appear in a system in question it would be worth investigating in depth what it implies. One can take the Poincar\'e and the Ruppeiner method on the same footing that it helps detect instabilities.}. Geometrical patterns of BH thermodynamics uncovered may play an important role in the context of quantum gravity.

\section*{Acknowledgments}

NP would like to acknowledge the organizers of the ERE2009 in Bilbao for their kind support, and he would also like to thank Roberto Emparan for useful discussions. NP and J{\AA} thank Ingemar Bengtsson on various comments. Finally NP warmly thanks the KoF group and Department of Physics, Stockholm University for the kind hospitality.

\end{document}